\documentclass{article}
\usepackage{amsmath, amssymb, amsthm, graphicx, hyperref, xcolor}
\usepackage{adjustbox}

\title{A Survey: Potential Dimensionality Reduction Methods For Data Reduction}
\author{Foo Hui Mean and Yuan-chin Ivan Chang\\
Institute of Statistical Science\\
Academia Sinica, Taipei, Taiwan 11529}
\date{\today}

\begin{document}

\maketitle

\begin{abstract}
Dimensionality reduction is a fundamental technique in machine learning and data analysis, enabling efficient representation and visualization of high-dimensional data. This paper explores five key methods: Principal Component Analysis (PCA), Kernel PCA (KPCA), Sparse Kernel PCA, t-Distributed Stochastic Neighbor Embedding (t-SNE), and Uniform Manifold Approximation and Projection (UMAP). PCA provides a linear approach to capturing variance, whereas KPCA and Sparse KPCA extend this concept to non-linear structures using kernel functions. Meanwhile, t-SNE and UMAP focus on preserving local relationships, making them effective for data visualization. Each method is examined in terms of its mathematical formulation, computational complexity, strengths, and limitations. The trade-offs between global structure preservation, computational efficiency, and interpretability are discussed to guide practitioners in selecting the appropriate technique based on their application needs.
\end{abstract}


\section{Introduction}

{
Dimensionality reduction techniques play a crucial role in analyzing and visualizing high-dimensional data by transforming it into a lower-dimensional space while preserving its essential structure. These methods help mitigate the curse of dimensionality, reduce computational costs, and reveal underlying patterns that may not be apparent in the original feature space.

This document provides an in-depth exploration of five key dimensionality reduction techniques: {\it Principal Component Analysis (PCA), Kernel PCA (KPCA), Sparse Kernel PCA, t-SNE, and UMAP}. Each method is discussed in terms of its mathematical foundation, computational complexity, advantages, and limitations. PCA serves as a fundamental linear approach, while KPCA and Sparse KPCA extend this idea to non-linear transformations using kernel methods. On the other hand, t-SNE and UMAP focus on preserving local relationships, making them particularly useful for visualization. By understanding the trade-offs between these techniques, practitioners can make informed decisions on which method best suits their data and application needs.
}

\section{Some Popular Dimension Deduction Methods}

\subsection{Principal Component Analysis (PCA)}

{
Principal Component Analysis (PCA) is a widely used dimensionality reduction technique that identifies the most important directions, called principal components, in a dataset \cite{Jolliffe2002}. These principal components are orthogonal directions that maximize variance, allowing us to reduce the number of dimensions while retaining as much information as possible.

Given a dataset $\mathbf{X} \in \mathbb{R}^{n \times d}$, PCA follows these steps:

\begin{enumerate}
    \item \textbf{Centering the Data:}  
    The mean of each feature is subtracted so that the dataset is centered at zero:
    \[
    \mathbf{X}_{\text{centered}} = \mathbf{X} - \bar{\mathbf{X}}
    \]

    \item \textbf{Computing the Covariance Matrix:}  
    The covariance matrix captures relationships between features:
    \[
    \mathbf{C} = \frac{1}{n} \mathbf{X}_{\text{centered}}^T \mathbf{X}_{\text{centered}}
    \]
    
    \item \textbf{Performing Eigen Decomposition:}  
    The covariance matrix is decomposed into eigenvalues and eigenvectors:
    \[
    \mathbf{C} = \mathbf{V} \Lambda \mathbf{V}^T
    \]
    The eigenvectors (principal components) represent the new coordinate axes, and their corresponding eigenvalues indicate the amount of variance they capture.

    \item \textbf{Selecting the Top $k$ Principal Components:}  
    The $k$ eigenvectors with the largest eigenvalues are chosen to form the reduced dataset. This allows us to project the original data onto a lower-dimensional space while preserving the most important patterns.
\end{enumerate}

PCA is highly effective for reducing dimensionality and simplifying data visualization. However, its limitations arise from its mathematical properties.

\paragraph{Limitations:}
\begin{itemize}
    \item \textbf{Assumes Linear Relationships:}  
    PCA works well when the data follows linear patterns but struggles with complex, non-linear structures.

    \item \textbf{Sensitive to Outliers:}  
    Since PCA is based on variance, extreme values (outliers) can distort the results and shift principal components in misleading directions.

    \item \textbf{Requires Normalization:}  
    PCA is affected by the scale of different features. If the data is not normalized (e.g., some features have much larger values than others), the principal components may be biased toward higher-magnitude features.

    \item \textbf{Loss of Interpretability:}  
    When too many principal components are retained, the transformed features may become difficult to interpret, as they are linear combinations of the original variables.
\end{itemize}

\textbf{Pros:} Fast, computationally efficient, preserves global structure, and provides an interpretable transformation in lower dimensions. \\
\textbf{Cons:} Assumes linear relationships, sensitive to outliers, and may require careful preprocessing (e.g., normalization).

}

\subsection{Kernel Principal Component Analysis (Kernel PCA)}

{
Kernel Principal Component Analysis (KPCA) extends traditional Principal Component Analysis (PCA) to capture nonlinear structures by leveraging the kernel trick \cite{Scholkopf1997}. Instead of performing eigen-decomposition on the original data covariance matrix, KPCA first maps data into a higher-dimensional feature space using a nonlinear function:
\[
\phi: \mathbb{R}^d \to \mathbb{R}^D, \quad D \gg d
\]
where $D$ is often infinite in certain kernel settings. The mapping $\phi(\cdot)$ is not computed explicitly; instead, the **kernel trick** allows us to compute inner products in the high-dimensional space using a kernel function:
\[
k(x_i, x_j) = \phi(x_i)^T \phi(x_j)
\]
Common kernel functions include the Radial Basis Function (RBF), polynomial, and sigmoid kernels. By replacing dot products with kernel evaluations, we effectively apply PCA in a high-dimensional space without explicitly computing the transformation.

The central computation in KPCA involves performing eigen-decomposition on the {\bf kernel matrix} $\mathbf{K}$:
\begin{equation}
    \mathbf{K} \mathbf{\alpha} = \lambda \mathbf{\alpha}
\end{equation}
where $\mathbf{K} \in \mathbb{R}^{n \times n}$ is the Gram matrix containing pairwise kernel evaluations, and $\mathbf{\alpha}$ are the eigenvectors corresponding to the principal components in the kernel space.

KPCA is particularly useful for discovering nonlinear patterns that standard PCA cannot capture. However, it also introduces several challenges, particularly in terms of computational complexity and stability.

\paragraph{Limitations:}
\begin{itemize}
    \item \textbf{High Computational Cost and Memory Usage:}  
    Since KPCA requires computing and storing the full kernel matrix $\mathbf{K} \in \mathbb{R}^{n \times n}$, it has a memory complexity of $O(n^2)$ and a computational complexity of $O(n^3)$ due to the eigen-decomposition step. This makes it impractical for large datasets.

    \item \textbf{Choice of Kernel Function is Crucial:}  
    The performance of KPCA depends heavily on selecting an appropriate kernel function. A poorly chosen kernel may fail to capture the underlying data structure or introduce distortions in the transformed space.

    \item \textbf{No Explicit Inverse Mapping:}  
    Unlike PCA, which allows direct reconstruction of the original data from the reduced representation, KPCA does not provide an explicit inverse transformation. Approximate pre-images can be computed, but these methods introduce errors.

    \item \textbf{Sensitivity to Kernel Hyperparameters:}  
    The effectiveness of KPCA is highly dependent on the hyperparameters of the chosen kernel. For instance, in the RBF kernel, the bandwidth parameter $\gamma$ controls how much influence each data point has, and in polynomial kernels, the degree $d$ affects the complexity of the decision boundary. Improper tuning can lead to underfitting or overfitting.
\end{itemize}

\textbf{Pros:} Effectively captures complex, nonlinear relationships in data, making it useful for pattern recognition and feature extraction in cases where PCA falls short. \\
\textbf{Cons:} Computationally expensive due to large matrix operations, requires careful selection of a kernel function and its parameters, and lacks an explicit inverse transformation.

}

\subsection{Sparse Kernel Principal Component Analysis (Sparse KPCA)}

{
Sparse Kernel Principal Component Analysis (Sparse KPCA) is an approximation of standard Kernel PCA that improves scalability by selecting a subset of representative training points \cite{Bach2002, Ong2003}. While Kernel PCA applies an eigendecomposition to a full Gram matrix $\mathbf{K} \in \mathbb{R}^{n \times n}$, Sparse KPCA reduces computational complexity by working with a significantly smaller Gram matrix constructed from a subset of $m$ representative points, where $m \ll n$.

The method follows a similar eigen-decomposition process:
\begin{equation}
    \mathbf{K}_m \mathbf{\alpha} = \lambda \mathbf{\alpha}, \quad \mathbf{K}_m \in \mathbb{R}^{m \times m}, \quad m \ll n
\end{equation}
where $\mathbf{K}_m$ is the reduced Gram matrix computed using only a subset of $m$ selected data points. These points act as a compressed representation of the full dataset, allowing for a computationally efficient approximation of Kernel PCA. This significantly reduces memory requirements and computational complexity, making KPCA feasible for larger datasets that would otherwise be intractable.

However, this approximation introduces certain trade-offs, particularly regarding accuracy and performance. The choice of the representative subset is critical, as it directly impacts the quality of the low-dimensional embedding.

\paragraph{Limitations:}
\begin{itemize}
    \item \textbf{Approximation Error:}  
    Since Sparse KPCA operates on a reduced Gram matrix rather than the full kernel matrix, it introduces an inherent approximation error. The extent of this error depends on how well the selected subset represents the overall data distribution.

    \item \textbf{Dependency on Support Vector Selection:}  
    The quality of the Sparse KPCA embedding is highly sensitive to the method used to select the subset of $m$ representative points. Poor selection may lead to suboptimal representations, reducing the effectiveness of the dimensionality reduction.

    \item \textbf{Computational Cost for Large Datasets:}  
    Although Sparse KPCA is more scalable than standard Kernel PCA, it still involves solving an eigenproblem of size $m \times m$, which can be computationally expensive when $m$ is large. Additionally, selecting the subset efficiently can introduce extra overhead.
\end{itemize}

\textbf{Pros:} Reduces computational cost compared to full Kernel PCA, makes kernel-based dimensionality reduction feasible for large datasets.\\
\textbf{Cons:} Introduces approximation errors, requires careful selection of representative points, and remains computationally intensive for very large datasets.

}

\subsection{t-Distributed Stochastic Neighbor Embedding (t-SNE)}

{
t-SNE \cite{van2008visualizing} (t-Distributed Stochastic Neighbor Embedding) is a non-linear dimensionality reduction technique designed primarily for high-dimensional data visualization. It models pairwise similarities in both high- and low-dimensional spaces using probability distributions, ensuring that similar data points remain close together in the embedded space.

The algorithm consists of the following key steps:
\begin{enumerate}
    \item \textbf{Constructing a Probability Distribution in the High-Dimensional Space:}  
    The first step involves computing a probability distribution $p_{ij}$ that represents pairwise similarities between data points in the original high-dimensional space. This is done using a Gaussian kernel, where the similarity between points $x_i$ and $x_j$ is determined by conditional probabilities based on a perplexity parameter.
    
    \item \textbf{Defining a Corresponding Probability Distribution in the Low-Dimensional Space:}  
    A similar probability distribution $q_{ij}$ is defined in the lower-dimensional space. Unlike the high-dimensional distribution, t-SNE uses a Student’s t-distribution with a single degree of freedom (equivalent to a Cauchy distribution), which has heavier tails. This helps prevent the "crowding problem" and ensures that dissimilar points are better separated.

    \item \textbf{Minimizing the KL Divergence Between $P$ and $Q$:}  
    The final step involves optimizing the lower-dimensional embeddings by minimizing the Kullback-Leibler (KL) divergence between $P$ and $Q$. KL divergence acts as a loss function, penalizing mismatches between the two distributions. Gradient descent is used to iteratively adjust the embedding until convergence.
\end{enumerate}

Since t-SNE is optimized for visualization rather than general-purpose dimensionality reduction, its embeddings do not preserve global distances or support tasks like classification or regression. Consequently, implementations such as the R package for t-SNE often restrict the number of output dimensions to 2 or 3.

\paragraph{Limitations:}
\begin{itemize}
    \item \textbf{Computational Complexity:}  
    t-SNE is computationally expensive, with a complexity of $O(n^2)$ due to the need to compute pairwise similarities. For large datasets, this makes it impractical. However, approximations such as Barnes-Hut t-SNE reduce complexity to $O(n \log n)$, making it more scalable.

    \item \textbf{Poor Global Structure Preservation:}  
    The method primarily preserves local neighborhoods but does not maintain global distances. As a result, distances between clusters in the 2D or 3D embedding space may not reflect true relationships in the original data.

    \item \textbf{Non-Deterministic Results:}  
    Even when fixing a random seed, t-SNE’s stochastic optimization process can still produce slightly different embeddings due to variations in early exaggeration and gradient descent behavior. 

    \item \textbf{Perplexity Sensitivity:}  
    The perplexity parameter, which controls the balance between local and global structure, must be carefully tuned for different datasets. Too low a perplexity may fragment clusters, while too high a value can obscure local structure.

    \item \textbf{Challenges with High-Dimensional Data:}  
    For very high-dimensional datasets ($d > 50$), t-SNE often benefits from an initial PCA step to reduce noise and improve distance calculations. Typically, PCA is used to bring the data down to 30–50 dimensions before applying t-SNE.
\end{itemize}

\textbf{Pros:} Excellent for visualizing clusters, effectively preserves local structure, widely used in exploratory data analysis. \\
\textbf{Cons:} Computationally expensive, non-deterministic, does not preserve global distances, requires careful tuning of perplexity.

}

\subsection{Uniform Manifold Approximation and Projection}

{
UMAP \cite{mcinnes2018umap} (Uniform Manifold Approximation and Projection) is a non-linear dimensionality reduction technique designed to preserve both local and some global structures of high-dimensional data while being computationally efficient. The method is built upon principles from fuzzy topology and Riemannian geometry, providing a theoretical framework for learning a lower-dimensional representation of data.

The algorithm works in two main steps:
\begin{enumerate}
\item \textbf{Graph Construction:} A weighted k-nearest neighbor (k-NN) graph is built from the high-dimensional data, where the edge weights represent the probability that points are connected based on a fuzzy simplicial set. This structure effectively captures local relationships between data points.
\item \textbf{Optimization via Low-Dimensional Embedding:} A lower-dimensional representation is learned by minimizing a cross-entropy loss that aligns the low-dimensional embedding with the fuzzy simplicial graph structure. The optimization process attempts to preserve both local connectivity and some degree of global structure.
\end{enumerate}
Unlike t-SNE, which explicitly models pairwise similarities, UMAP constructs a global topology of the data before optimizing the embedding, making it more effective for preserving large-scale structures.

\paragraph{Limitations:}
\begin{itemize}
\item \textbf{Sensitivity to Hyperparameters:} The quality of the embeddings is highly dependent on parameters such as $n_neighbors$ (which controls the trade-off between local and global structure) and $min_dist$ (which regulates the minimum spacing between points in the low-dimensional space). Choosing inappropriate values may lead to distorted embeddings.
\item \textbf{No Explicit Inverse Transformation:} Unlike PCA or autoencoders, UMAP does not provide a direct mathematical function to map reduced points back to the original space, making reconstruction difficult. However, approximate inverse mappings can be learned with additional models.
\item \textbf{Non-Deterministic Results:} The stochastic nature of the optimization process means that different runs may produce slightly different embeddings unless a fixed random seed is used. Although more stable than t-SNE, results can still vary depending on initialization.
\item \textbf{Limited Global Structure Preservation:} While UMAP retains more global structure than t-SNE, it is still primarily optimized for local relationships. Large-scale distances may not always be faithfully preserved, especially when reducing to very low dimensions (e.g., 2D or 3D).
\item \textbf{Preprocessing for High-Dimensional Data:} For very high-dimensional datasets ($d > 100$), UMAP often benefits from an initial PCA step to reduce noise and improve embedding stability, as raw high-dimensional distances can be unreliable.
\end{itemize}

\textbf{Pros:} Faster than t-SNE, scalable to large datasets, effectively preserves local structure while maintaining some global relationships. \
\textbf{Cons:} Sensitive to hyperparameters, non-deterministic without a fixed seed, lacks an explicit inverse transformation.

}

\subsection{Comparison of Methods}
Table \ref{tab:comparison} provides a comparative overview of dimensionality reduction techniques, highlighting their ability to handle high-dimensional data ($d$), large sample sizes ($n$), preservation of global structure, and computational complexity.

PCA is an efficient and interpretable method that scales well with large datasets but is limited to linear structures. Kernel PCA (K-PCA) extends PCA to non-linear data but suffers from high computational costs and the curse of dimensionality. Sparse Kernel PCA (SK-PCA) mitigates K-PCA’s scalability issues by selecting a subset of representative points, reducing complexity while introducing approximation errors.

For visualization, t-SNE effectively captures local structures but is computationally expensive and distorts global relationships. UMAP improves upon t-SNE by being significantly faster ($O(n \log n)$) and preserving more global structure, though it remains sensitive to hyperparameter choices.

The choice of method depends on dataset characteristics and computational constraints. PCA and UMAP are preferable for large datasets, while K-PCA and t-SNE are suited for capturing non-linearity and local patterns in smaller datasets.

\begin{table}[h]
\caption{Comparison of Dimensionality Reduction Methods}
       \label{tab:comparison}
    \centering
    \begin{adjustbox}{max width=\textwidth}
    \begin{tabular}{|l|c|c|c|c|}
        \hline
        \textbf{Method} & \textbf{High $d$?} & \textbf{Large $n$?} & \textbf{Global Structure?} & \textbf{Complexity} \\
        \hline
        PCA & \checkmark (but linear) & \checkmark & \checkmark & $O(nd^2)$ \\
         K-PCA & $\times$ (Curse of Dimensionality) & $\times$ & \checkmark & $O(n^3)$ \\
        SK-PCA & \checkmark & $\sim$ (Approximation Error) & \checkmark & $O(m^3)$ where $m \ll n$ \\
        t-SNE & $\times$ ($d > 50$ problematic) & $\times$ ($O(n^2)$) & $\times$ & $O(n^2)$ \\
        UMAP & \checkmark & \checkmark ($O(n \log n)$) & $\times$ (better than t-SNE) & $O(n \log n)$ \\
        \hline
    \end{tabular}
    \end{adjustbox}
         d: number of dimensionality; n: sample size; K-PCA:Kernel PCA; SK-PCA:Sparse Kernel PCA; 
\end{table}

\section{Conclusion}
Each dimensionality reduction method involves trade-offs between preserving global and local structure, computational efficiency, and interpretability. Principal Component Analysis (PCA) is a computationally efficient and interpretable technique that effectively preserves global variance but struggles with non-linear data. Kernel PCA (KPCA) extends PCA to capture non-linear relationships through the use of kernel functions, though it is computationally expensive and requires careful selection of an appropriate kernel. Sparse Kernel PCA improves scalability over KPCA by reducing the dataset to a subset of representative points, but this introduces approximation errors that can affect accuracy.

Among non-linear methods focused on visualization, t-SNE is highly effective in capturing local structures but tends to distort global distances and comes with a high computational cost. In contrast, UMAP provides a more scalable and efficient alternative to t-SNE while preserving more global structure. However, it remains sensitive to hyperparameter choices, requiring careful tuning for different datasets.

The optimal choice of dimensionality reduction method depends on the characteristics of the dataset, the need for interpretability, computational constraints, and the specific goals of the analysis. PCA is best suited for preserving global variance, KPCA for modeling complex non-linear relationships, and Sparse KPCA for balancing scalability with kernel-based transformations. t-SNE is ideal when local structure is the primary concern, whereas UMAP provides a practical compromise between computational efficiency and structure preservation. Ultimately, selecting the right technique requires a balance between domain knowledge, familiarity with statistical and mathematical tools, and an understanding of the trade-offs each method presents.


\begin{thebibliography}{9}
    \bibitem{Jolliffe2002} I. T. Jolliffe, \textit{Principal Component Analysis}, Springer, 2002.
    \bibitem{Scholkopf1997} B. Schölkopf, A. Smola, and K.-R. Müller, \textit{Kernel Principal Component Analysis}, Neural Computation, 1997.
    \bibitem{Bach2002} F. R. Bach and M. I. Jordan, \textit{Kernel Independent Component Analysis}, Journal of Machine Learning Research, 2002.
    \bibitem{Ong2003} C. S. Ong, A. J. Smola, and R. C. Williamson, \textit{Learning the Kernel with Hyperkernels}, Journal of Machine Learning Research, 2003.
    \bibitem{van2008visualizing} L. van der Maaten and G. Hinton, \textit{Visualizing Data using t-SNE}, Journal of Machine Learning Research, 2008.
    \bibitem{mcinnes2018umap} L. McInnes, J. Healy, and J. Melville, \textit{UMAP: Uniform Manifold Approximation and Projection for Dimension Reduction}, arXiv:1802.03426, 2018.
\end{thebibliography}
\end{document}